Methodology

# Evaluation of cell wall preparations for proteomics: a new procedure for purifying cell walls from Arabidopsis hypocotyls

Leila Feiz†[1,2], Muhammad Irshad†[1], Rafael F Pont-Lezica[1], Hervé Canut[1] and Elisabeth Jamet*[1]

Address: [1]Surfaces Cellulaires et Signalisation chez les Végétaux, UMR 5546 CNRS-Université Paul Sabatier-Toulouse III, Pôle de Biotechnologie végétale, 24, Chemin de Borde Rouge, BP 42617 Auzeville, 31326 Castanet-Tolosan, France and [2]Department of Plant Science and Plant Pathology, Agriculture and Biological Sciences Faculty, Montana State University-Bozeman, Bozeman, MT 59717-3150, USA

Email: Leila Feiz - lfeiz@mymail.msu.montana.edu; Muhammad Irshad - muhammad@scsv.ups-tlse.fr; Rafael F Pont-Lezica - lezica@scsv.ups-tlse.fr; Hervé Canut - canut@scsv.ups-tlse.fr; Elisabeth Jamet* - jamet@scsv.ups-tlse.fr

* Corresponding author †Equal contributors

## Abstract

**Background:** The ultimate goal of proteomic analysis of a cell compartment should be the exhaustive identification of resident proteins; excluding proteins from other cell compartments. Reaching such a goal closely depends on the reliability of the isolation procedure for the cell compartment of interest. Plant cell walls possess specific difficulties: (i) the lack of a surrounding membrane may result in the loss of cell wall proteins (CWP) during the isolation procedure, (ii) polysaccharide networks of cellulose, hemicelluloses and pectins form potential traps for contaminants such as intracellular proteins. Several reported procedures to isolate cell walls for proteomic analyses led to the isolation of a high proportion (more than 50%) of predicted intracellular proteins. Since isolated cell walls should hold secreted proteins, one can imagine alternative procedures to prepare cell walls containing a lower proportion of contaminant proteins.

**Results:** The rationales of several published procedures to isolate cell walls for proteomics were analyzed, with regard to the bioinformatic-predicted subcellular localization of the identified proteins. Critical steps were revealed: (i) homogenization in low ionic strength acid buffer to retain CWP, (ii) purification through increasing density cushions, (iii) extensive washes with a low ionic strength acid buffer to retain CWP while removing as many cytosolic proteins as possible, and (iv) absence of detergents. A new procedure was developed to prepare cell walls from etiolated hypocotyls of *Arabidopsis thaliana*. After salt extraction, a high proportion of proteins predicted to be secreted was released (73%), belonging to the same functional classes as proteins identified using previously described protocols. Finally, removal of intracellular proteins was obtained using detergents, but their amount represented less than 3% in mass of the total protein extract, based on protein quantification.

**Conclusion:** The new cell wall preparation described in this paper gives the lowest proportion of proteins predicted to be intracellular when compared to available protocols. The application of its principles should lead to a more realistic view of the cell wall proteome, at least for the weakly bound CWP extractable by salts. In addition, it offers a clean cell wall preparation for subsequent extraction of strongly bound CWP.

## Background

Cell walls are natural composite structures, mostly made of high molecular weight polysaccharides, proteins, and lignins, the latter found only in specific cell types. They are dynamic structures contributing to the general morphology of the plant. Cell walls are involved in cell expansion and division, and they are sources of signals for molecular recognition within the same or between different organisms [1-5]. Cell wall proteins (CWP) represent a minor fraction of the wall mass: 5–10% in primary cell walls of dicots, as reported for cell suspension cultures, but accurate determinations in various plant organs are still lacking [6]. Despite their low abundance, CWP contribute, at least in part, to the dynamic of cell walls. CWP can be involved in modification of cell wall components, wall structure, signalling, and interactions with plasma membrane proteins at the cell surface [7].

Proteomics appears to be a suitable method to identify a large number of CWP thus providing information for many genes still lacking a function. Recent publications on cell wall proteomics have shown that more than 50% of the identified proteins were known to be intracellular proteins in higher plants [8,9], green alga [10] and fungi [11]. Different techniques unrelated to proteomics, such as biotinylation of cell surface proteins, or immunoelectron microscopy, also suggested a cell wall location for some glycolytic enzymes, proposing that they are *bona fide* components of the yeast cell wall [11]. However, the reliability of protein profiling for a compartment like the cell wall, strongly depends on the quality of the preparation. Unfortunately, the classical methods to check the purity of a particular fraction are not conclusive for proteomic studies, since the sensibility of the analysis by mass spectrometry is 10 to 1000 times more sensitive than enzymatic or immunological tests using specific markers. Our experience in the field has shown that the most efficient way to evaluate the quality of a cell wall preparation is (i) to identify all the proteins extracted from the cell wall by mass spectrometry, and (ii) to perform extensive bioinformatic analysis to determine if the identified proteins contain a signal peptide, and no retention signals for other cell compartments [12-15]. It is then possible to conclude about the quality of the cell wall preparation by calculating the proportion of predicted secreted proteins to intracellular ones.

The aim of the present study is to present a comparative analysis of different methods previously published to prepare cell walls for proteomic studies. These methods will be evaluated by the proportion of proteins predicted to be secreted after bioinformatic analysis as stated above. A new method is presented, based on classical cell wall preparations, but adapted to the new technologies. The results indicate that such a method significantly reduces the number of proteins without predicted signal peptide.

## Results and discussion

Several strategies have been designed to gain access to CWP. The most labile CWP, *i.e.* those having little or no interactions with cell wall components, can be recovered in culture media of cell suspension cultures [12] or liquid cultured seedlings [14]. Extracellular fluids can be harvested from cell suspension cultures [12,16] or intact organs such as leaves [13]. However, such analysis cannot be done in all cases. It is then necessary to isolate cell walls starting with a drastic mechanical disruption of the material of interest. Consequently, labile CWP may be lost, and intracellular proteins or organelle fragments may contaminate cell wall preparations.

To design a procedure for cell wall isolation and subsequent protein extraction, several general features should be kept in mind. Plant and fungal cell walls are mainly built up with highly dense polysaccharides. This property can be used to purify them through density gradients by centrifugation. The biochemical structure of walls is complex, and CWP can be bound to the matrix by Van der Waals forces, hydrogen bonds, and hydrophobic or ionic attractions. Such interactions can also be modulated by the composition of the isolation medium. Commonly, a low ionic strength is preferred to preserve ionic bonds, but also to dilute the ionic strength of the cell wall itself. An acidic pH is chosen to maintain the interactions between proteins and polysaccharides as *in planta*. Once isolated, cell walls are classically treated by $CaCl_2$ buffers to release proteins, and by LiCl buffers for extraction of glycoprotein [17,18]. The use of detergents has also been reported to extract proteins strongly embedded in the polysaccharide matrix, like wall associated kinases [19]. Finally, CWP can be covalently bound to cell wall components so that they are resistant to salt-extraction. At present, there is no satisfactory procedure to isolate them. We analyze recent publications using different methods to isolate cell walls from plants or yeast prior to proteomic analysis [8,9,20].

### *Analysis of early methods*

Chivasa et al. [8] used *A. thaliana* cell cultures to purify cell walls. The procedure is schematically represented in Figure 1A. The first step consisted in cell grinding in water. The homogenate was layered onto 10% glycerol and let to sediment for several hours. The cell wall pellet was resuspended in water and washed 3 times by repeated centrifugations. The proteins were sequentially extracted with 0.2 M $CaCl_2$, and urea buffer (7 M urea, 2 M thiourea, 4% CHAPS, 1% DTT, 2% Pharmalytes 3–10). The extracted proteins were separated by 2D-GE and identified by MALDI-TOF mass spectrometry (MS). The identified proteins were analyzed with several bioinformatic programs

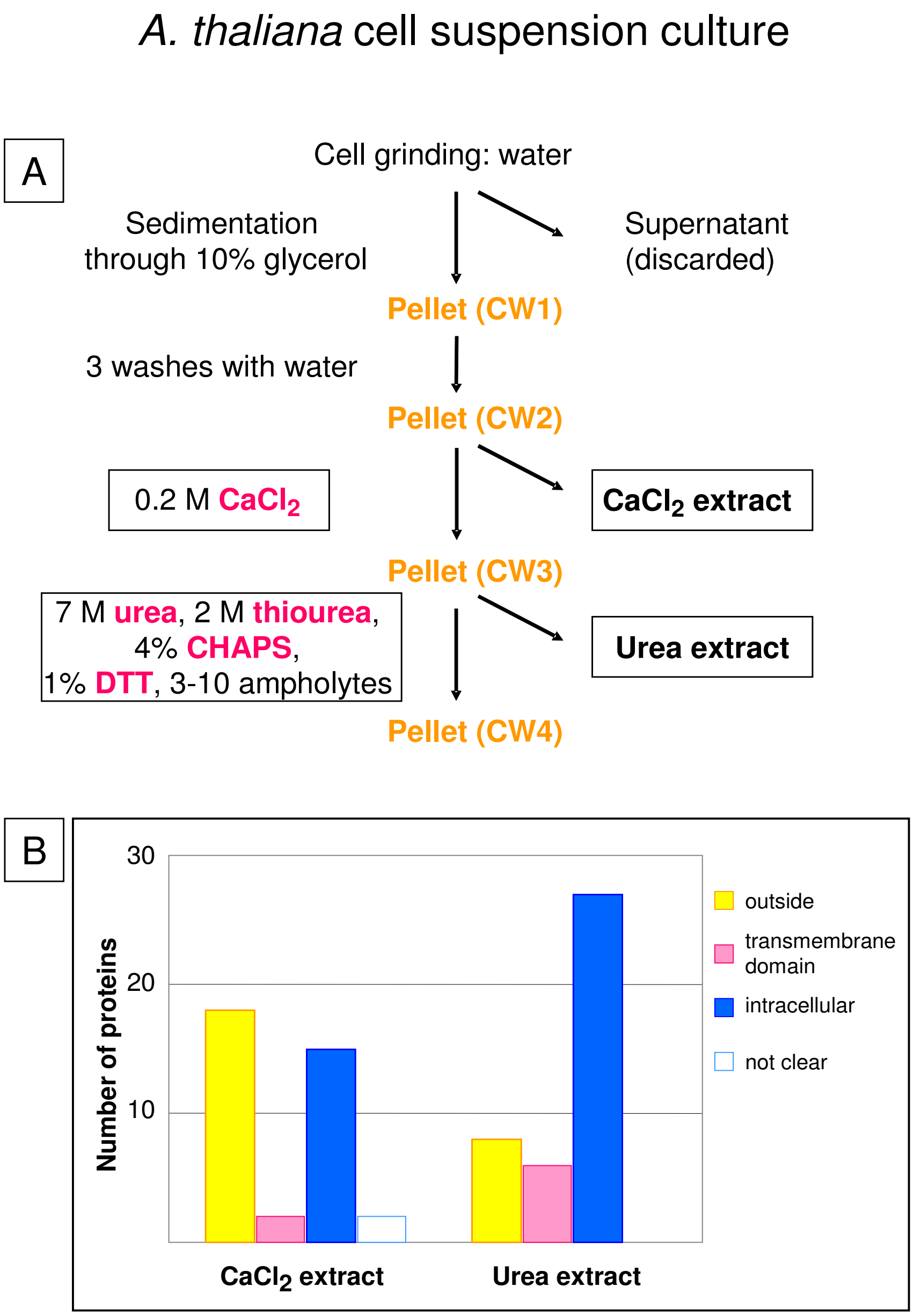

**Figure 1**
**Cell wall preparation from *A. thaliana* cell cultures. A –** Schematic representation of the purification of cell walls from *A. thaliana* cell suspension cultures, and of the different extracts obtained [8]. **B –** Number of proteins identified in each extract after separation by 2D-GE and MALDI-TOF MS analysis. After bioinformatic analysis, proteins were classified as outside (proteins containing a signal peptide and no other targeting sequences), having at least one trans-membrane domain and intracellular (predicted to be located in any intracellular compartment). Proteins for which predictions by different bioinformatic programs are in conflict are classified as "not clear". Twenty-four different proteins predicted to be secreted were identified in this study (see Additional file 1).

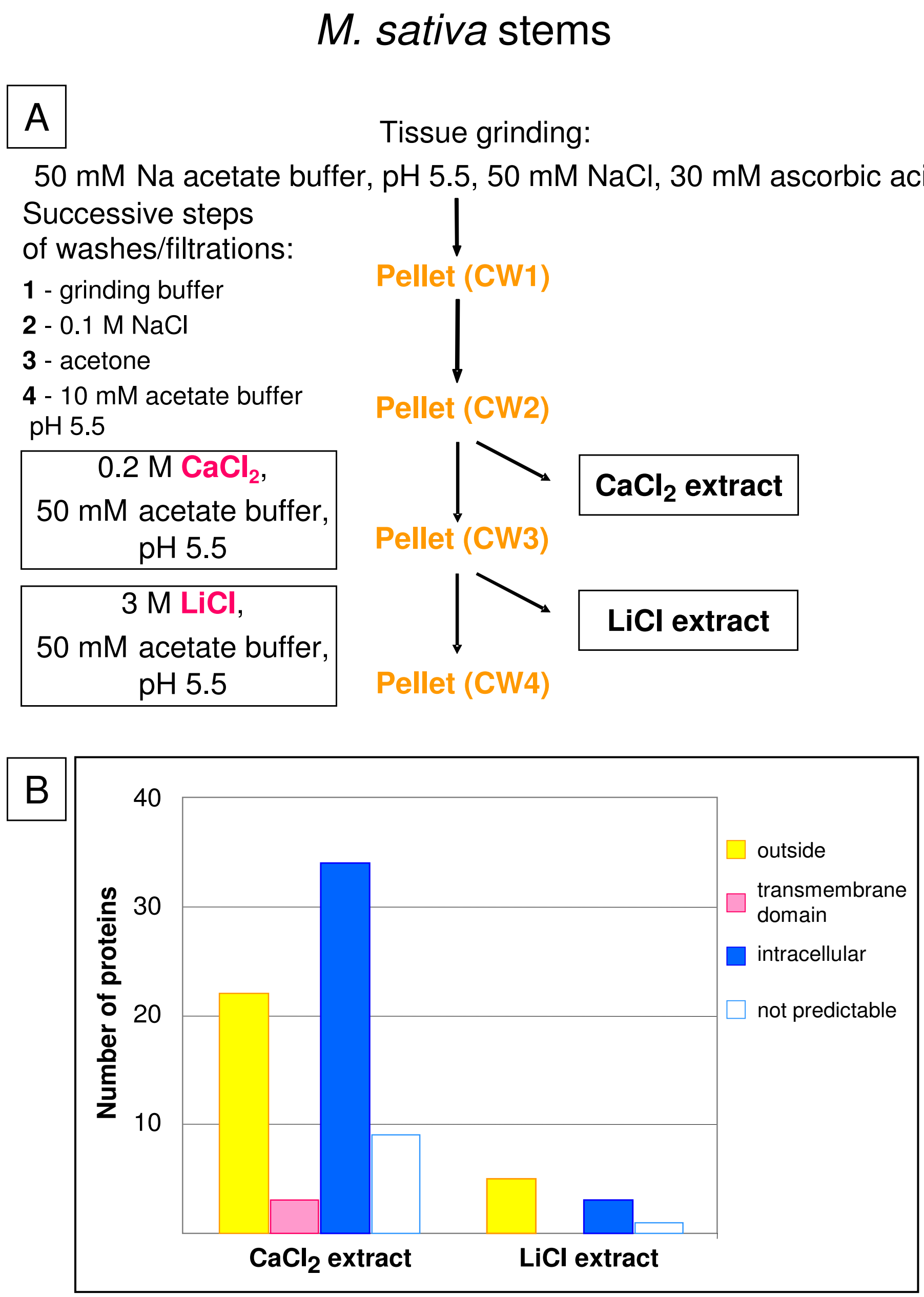

**Figure 2**
**Cell wall preparation from *M. sativa* stems. A –** Schematic representation of the purification procedure of cell walls from *M. sativa* stems, and of the different extracts obtained [9]. **B –** Number of proteins identified in each extract after 2D-GE separation and LC-MS/MS analysis. Proteins were classified as indicated in legend to Figure 1. Since the *M. sativa* genome is not fully sequenced, the sequence of the N-terminus of some proteins is not known. They were classified as "not predictable". Twenty-five different proteins predicted to be secreted were identified in this study (see Additional file 2).

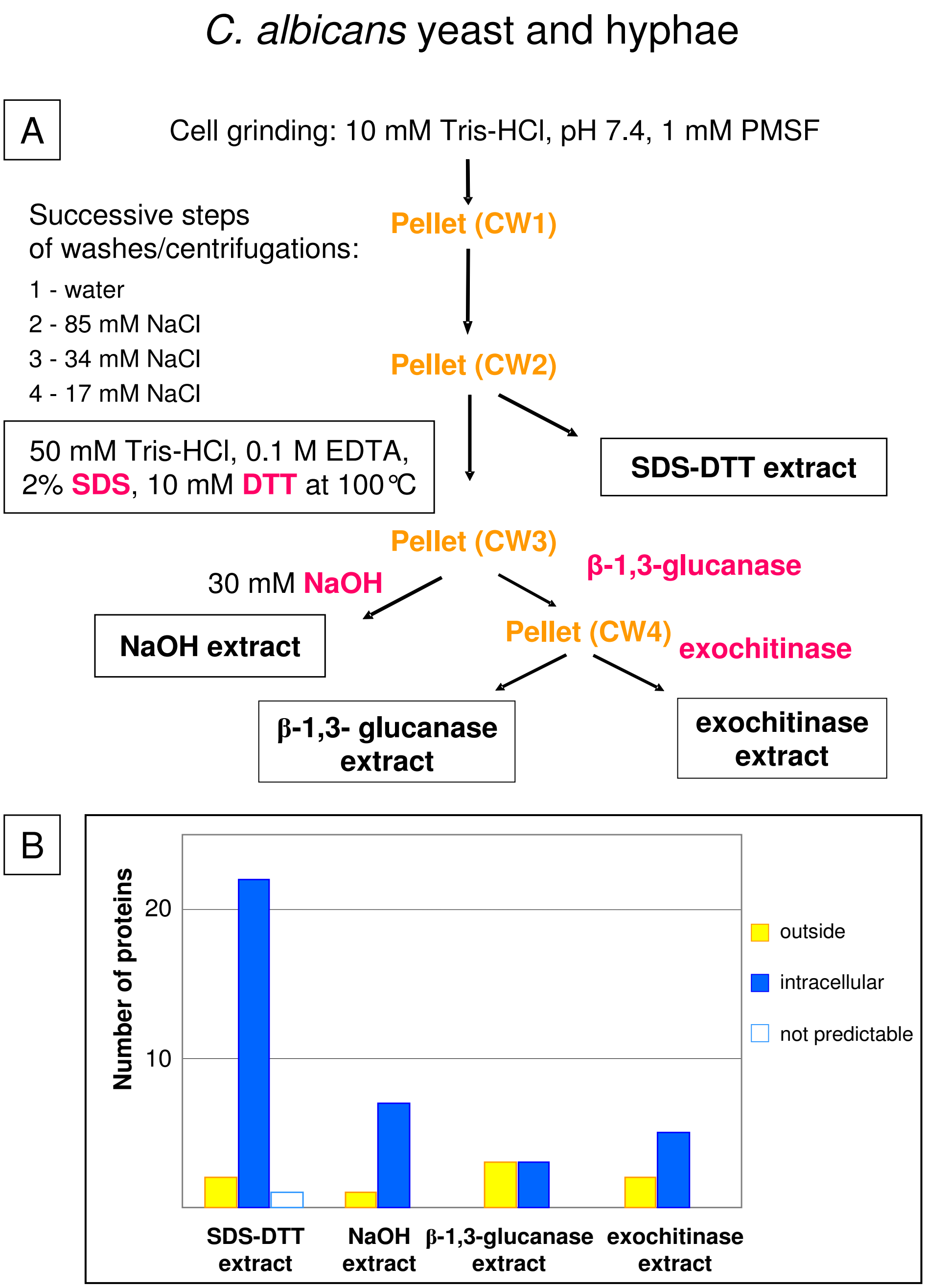

**Figure 3**

**Preparation of cell walls from *C. albicans* (yeast and hyphae). A –** Schematic representation of the purification procedure of cell walls from the dimorphic fungus *C. albicans* (yeast and hyphae), and of the extracts obtained [20]. **B –** Number of proteins identified in each extract after 2D-GE and analysis by MALDI-TOF or MALDI-TOF/TOF MS. Proteins were classified as indicated in legend to Figure 1. Four different proteins predicted to be secreted were identified in this study (see Additional file 3).

and classified as: (i) outside (proteins predicted to be secreted since they contain a signal peptide and no other targeting sequence), (ii) having trans-membrane domain(s), or (iii) intracellular (proteins not fulfilling these criteria). Figure 1B represents the results of both extractions. It appears that the $CaCl_2$ extract contains the highest proportion of proteins predicted to be secreted (50%), and that the use of detergents and chaotropic agents brings out mostly intracellular proteins, even if 20% of this fraction corresponds to predicted secreted proteins. Twenty-four different proteins predicted to be secreted were identified with this method.

Stems of *Medicago sativa* (alfalfa) were used for cell wall protein profiling [9]. In this case, a different procedure was used to isolate cell walls (Figure 2A). Tissues were frozen and ground in cold grinding buffer (50 mM Na acetate buffer, pH 5.5, 50 mM NaCl, and 30 mM ascorbic acid) with PVPP. Cell walls were isolated by filtering through a 47 $\mu m^2$ mesh nylon membrane and washed sequentially with grinding buffer, 0.1 M NaCl, acetone and 10 mM Na acetate pH 5.5. The proteins were sequentially extracted with 0.2 M $CaCl_2$, and 3 M LiCl buffers. The data obtained in this publication was analyzed as mentioned above, and the results are presented in Figure 2B. The proportion of intracellular proteins in the $CaCl_2$ extract is quite high (50%). It seems that the first washes do not eliminate such proteins. It is also possible that the wash performed with 0.1 M NaCl eliminates part of the secreted proteins. Twenty-five different proteins predicted to be secreted were identified in this study.

The procedure for isolation of cell walls from the dimorphic fungus *Candida albicans* used by Pitarch et al [20] was based on previous methods designed to isolate proteins covalently linked to the polysaccharide matrix [21,22]. Yeast and hyphae were collected by centrifugation and filtration, washed several times with lysis buffer (10 mM Tris-HCl, pH 7.4, 1 mM PMSF), and mechanically disrupted in lysis buffer (Figure 3A). After centrifugation, the pellet was successively washed with cold water and decreasing concentrations of NaCl (85, 34 and 17 mM) in 1 mM PMSF. Proteins were extracted with boiling SDS-extraction buffer (50 mM Tris-HCl, pH 8.0, 0.1 M EDTA, 2% SDS, 10 mM DTT). The residue was separated in two fractions, one was extracted with alkali (30 mM NaOH), and the other was submitted to sequential digestions by a β-1,3-glucanase followed by an exochitinase to break down the polysaccharide matrix. Each of the four samples were separated by 2D-GE, digested with trypsin and the peptides identified by MALDI-TOF or MALDI-TOF/TOF MS. The proteins identified in this publication were submitted to bioinformatic analysis and the results are represented in Figure 3B. Only four proteins predicted to be secreted were identified. All the others are predicted to be intracellular proteins (78%).

Altogether, this evaluation of three procedures to isolate cell walls from plants or fungi prior to proteomic analyses shows that they all produce a material containing a high proportion of proteins predicted to be intracellular, suggesting they are contaminants. Even if a careful bioinformatic analysis allows the discrimination between secreted and intracellular proteins, the time and effort consumed is not satisfactory.

### *A modified method to prepare plant cell walls*

From the analysis of the presented methods and other classical cell wall preparations used for the purification of cell wall enzymes [23,24], several points appear to be essential for the purification of cell walls. First, the presence of NaCl at early steps of cell wall preparations of *M. sativa* and *C. albicans* in grinding or washing buffers might induce a release of CWP even at a low concentration [12]. This might indirectly increase the proportion of intracellular proteins sticking non-specifically to cell wall polysaccharides. The use of a low ionic strength buffer for tissue grinding and subsequent washes to purify cell walls appears as an interesting alternative to prevent loss of CWP. Second, the protocol used for *A. thaliana* cell suspension cultures is the only one including a purification of cell walls through a dense medium, *i.e.* sedimentation in 10% glycerol. Proteins predicted to be secreted represented 50% of the identified proteins. It seems that a series of sedimentations/centrifugations in solutions of increasing densities would help in eliminating organelles and other vesicles less dense than cell wall polysaccharides [23,24]. Third, despite repeated washes, the cell wall preparation from *A. thaliana* cell suspension cultures still contained a high proportion of proteins predicted to be intracellular. A way to eliminate soluble contaminants such as intracellular proteins is to perform extensive washes of cell walls with a low ionic strength buffer [24]. Finally, the addition of polyvinyl polypyrrolidone (PVPP) to trap plant phenolic compounds [14] as well as anti-proteases to limit protein degradation during the manipulations [13], improves the quality of protein identification by mass spectrometry. Our procedure was established on the basis of these requirements.

Eleven day-old etiolated hypocotyls were ground in a low ionic strength buffer, 5 mM acetate buffer pH 4.6 in 0.4 M sucrose (Figure 4). PVPP and anti-proteases were added to the homogenate, centrifuged, and the resulting pellet resuspended in 5 mM acetate buffer pH 4.6 with increasing concentrations of sucrose (0.6 M and 1 M) and centrifuged. The residue (CW3) was extensively washed on a nylon membrane (25 μm pore size) with large amounts of 5 mM acetate buffer pH 4.6 (3 L for 16 g fresh material).

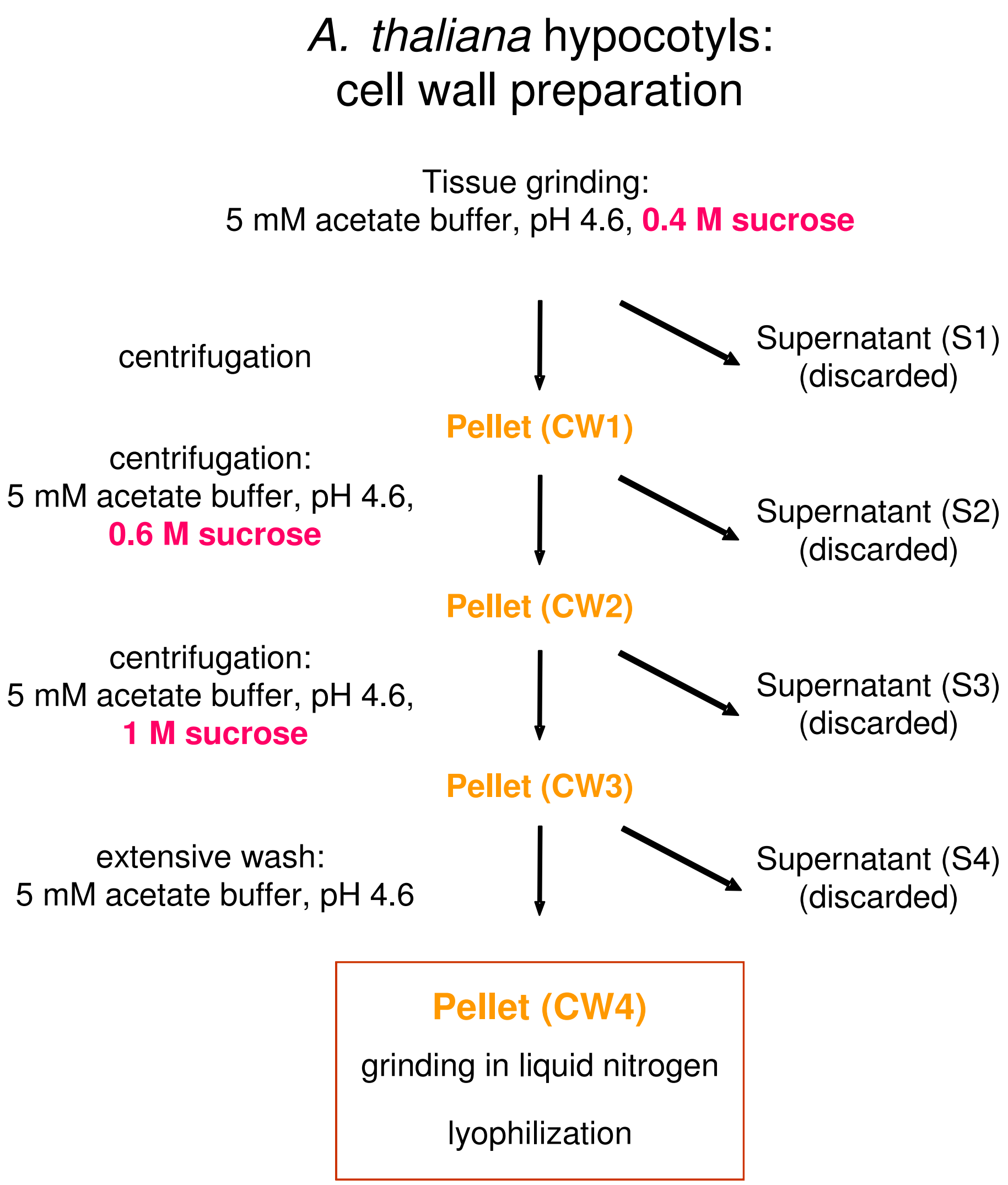

**Figure 4**
**Cell wall preparation from *A. thaliana* hypocotyls.** Schematic representation of the purification procedure of cell walls of *A. thaliana* etiolated hypocotyls. Sixteen g of fresh etiolated hypocotyls were ground. All supernatants were discarded after each centrifugation. The CW3 residue was extensively washed on a nylon net (25 µm pore size) with 3 L of 5 mM acetate buffer, pH 4.6. The CW4 pellet was ground in liquid nitrogen in a mortar with a pestle in order to reduce the size of the fragments, and lyophilized, obtaining 1.3 g of dry powder.

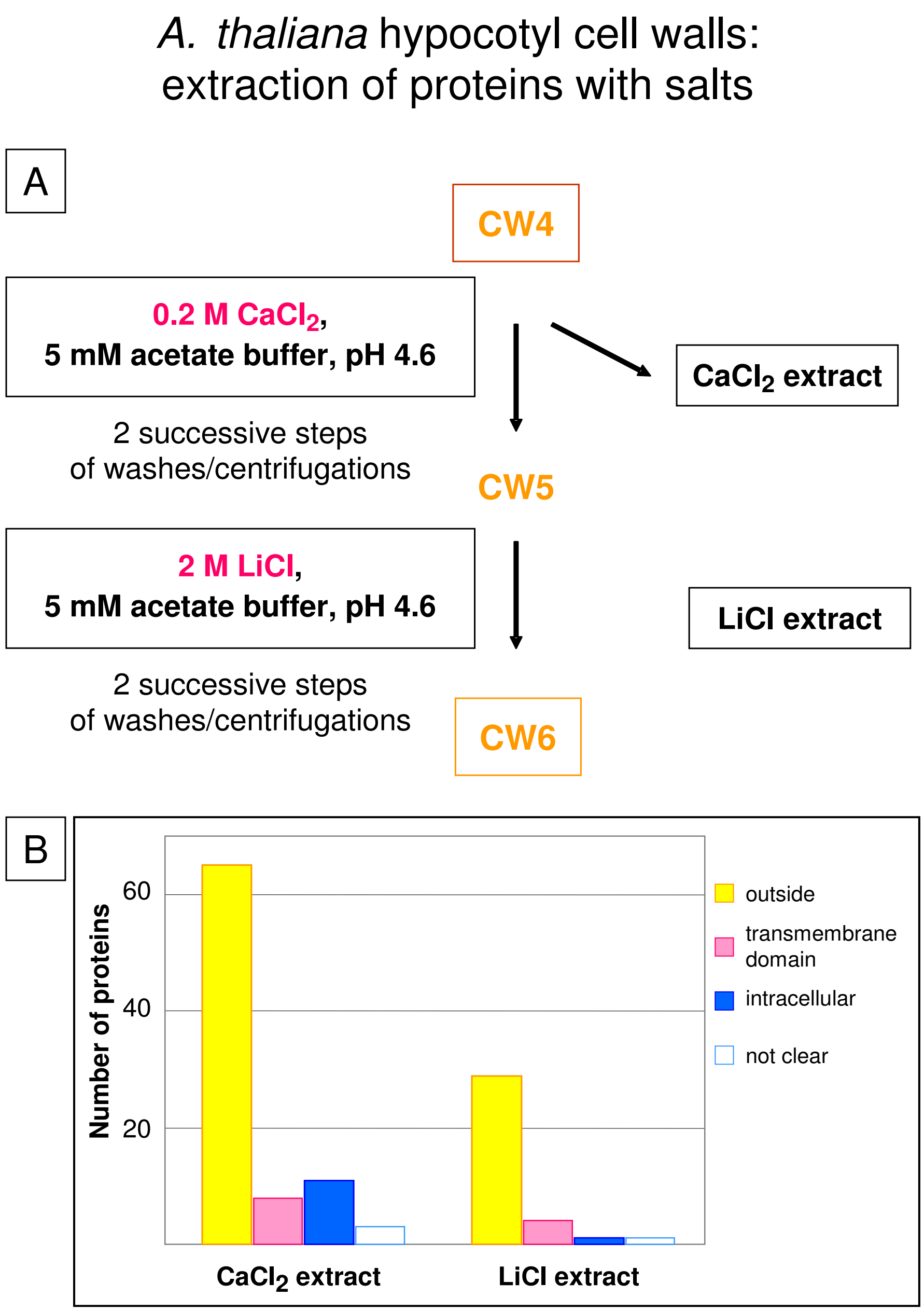

**Figure 5**

**Extraction of proteins from *A. thaliana* hypocotyls with salts. A –** Half of the CW4 lyophilized powder (0.65 g) was successively suspended in 0.2 M $CaCl_2$, 5 mM acetate buffer pH 4.6, and in 2 M LiCl, 5 mM acetate buffer pH 4.6. The $CaCl_2$ extract contained 400 µg of proteins. The LiCl extract contained 40 µg of proteins. **B –** Number of proteins identified in each extract after 1D-GE separation, and analysis by MALDI-TOF MS or LC-MS/MS. Proteins were classified as indicated in legend to Figure 1. Seventy-three different proteins predicted to be secreted were identified in this study (see Additional file 4).

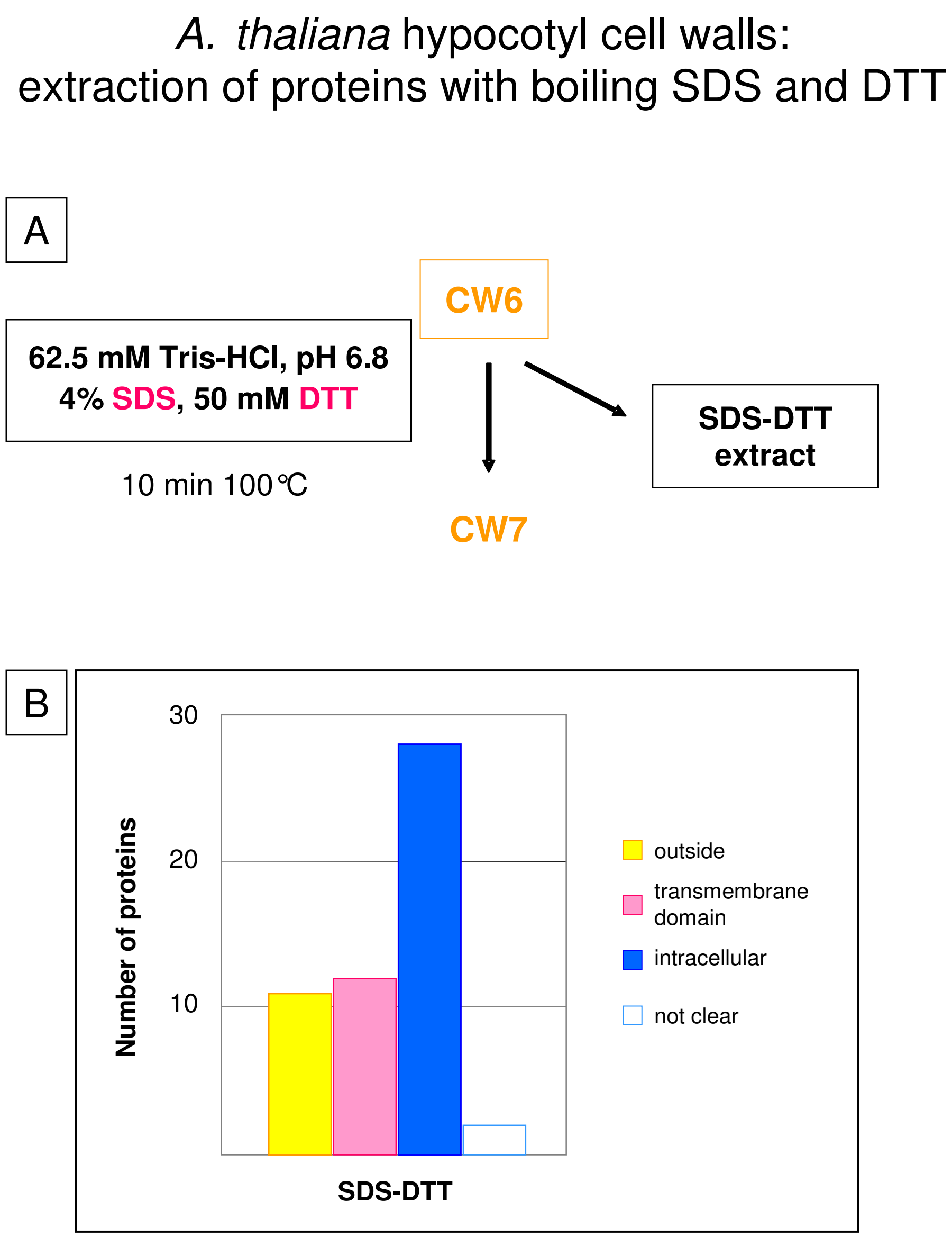

**Figure 6**
**Extraction of proteins from *A. thaliana* hypocotyls cell walls with boiling SDS and DTT. A –** The CW6 pellet described in Figure 5 was used for further extraction of proteins using boiling 4% SDS and 50 mM DTT in 62.5 mM Tris-HCl buffer, pH 6.8. The SDS-DTT extract contained 10 μg of proteins. **B –** Number of proteins identified in each extract after 1D-GE separation, and analysis by MALDI-TOF MS or LC-MS/MS. Proteins were classified as indicated in legend to Figure 1 (see Additional file 5).

The cell wall fraction was ground in liquid nitrogen in a mortar to reduce the size of the particles, which improves the subsequent extraction of proteins (CW4). In a typical cell wall preparation from *A. thaliana* hypocotyls, 16 g of fresh material resulted in 1.3 g of lyophilized powder.

### *Sequential salt extraction of proteins from cell walls*

The lyophilized powder (CW4) was sequentially extracted with salts that can extract proteins ionically bound to the polysaccharide matrix. Calcium chloride has been reported as an efficient salt for the extraction of cell wall proteins [8,9,13], and was used in this procedure (Figure 5A) to identify 65 secreted proteins (see Additional file 4). From 0.65 g of lyophilized powder, 400 μg of proteins were routinely obtained. LiCl was known as potent salt to extract hydroxyproline-rich glycoproteins from *Chlamydomonas reinhardii* cells, and was successfully used on *M. sativa* cell walls and *A. thaliana* rosettes [9,13,18]. LiCl only extracted 40 μg of proteins, 29 were identified as predicted to be secreted, but only 8 were specifically extracted by this salt (see Additional file 4).

A recent review on 281 CWP identified in proteomic studies by mass spectrometry [7] concluded that more than 60% of them have a basic pI, and around 80% in salt-extractable fractions. This is a serious problem for the separation of CWP by the classical 2D-GE, since it is well known that basic glycoproteins are poorly resolved by this technique [7]. We have then used 1D-GE for the separation, each band stained with Coomassie™ blue was digested by trypsin and further analyzed by MALDI-TOF MS or LC/MS/MS. Each protein was analyzed using several bioinformatic programs as described above. Seventy-three different proteins predicted to be secreted were identified in this study, whereas only 12 proteins predicted to be intracellular and 11 proteins predicted to have trans-membrane domains were found.

The protocol of CWP extraction we used is almost the same as the one employed with *M. sativa* stems [9]. But comparison of Figures 2B and 5B shows big differences in the proportion of proteins predicted to be intracellular or having trans-membrane domains (50% for *M. sativa vs* 27% for *A. thaliana*). Since the main difference between the two protocols is the addition of centrifugations through a dense medium, it shows that this step is critical for the purity of cell walls.

The proteins predicted to be secreted identified in this study belong to the same functional classes as those described in previous cell wall proteomes established from cell wall preparations [8,9,25]. Shortly, these functional classes comprise proteins acting on polysaccharides (*e.g.* glycoside hydrolases, carbohydrate esterases, expansins), proteases, proteins with interacting domains (*e.g.* lectins, leucine-rich repeat proteins, enzyme inhibitors), oxido-reductases (*e.g.* peroxidases, berberine-bridge enzymes), proteins involved in signaling processes (*e.g.* arabinogalactan proteins), structural proteins, proteins of unknown function and miscellaneous proteins [7]. The new protocol appears to be more efficient since a large proportion of identified proteins are predicted to be targeted to the compartment of interest.

Since we noticed that the use of detergents described in previous protocols increased the proportion of proteins predicted to be intracellular or having trans-membrane domains [8,20], we wanted to test it on our cell wall preparation. The CW6 pellet (Figure 6) was treated with boiling SDS-DTT buffer (62.5 M Tris-HCl, pH 6.8, 4% SDS, 50 mM DTT). Less than 10 μg of proteins were obtained. As for the other fractions, the proteins were concentrated, separated by 1D-GE, and the Coomassie™ blue stained bands were analyzed by mass spectrometry. Fifty-three proteins were identified in this fraction, among which 11 (20%) were predicted to be secreted, 12 (23%) were predicted to have trans-membrane domains and 30 (57%) were predicted to originate from intracellular compartments including cytoplasm, endoplasmic reticulum, microbodies, and chloroplasts. Comparison of these results with those shown in Figures 1B and 3B, in which detergents such as CHAPS-Urea-DTT or SDS-DTT were also used, confirms that treatment with detergents mainly extract intracellular proteins as well as membrane proteins trapped within the polysaccharide matrix: respectively 66% and 88% of proteins predicted to be intracellular, and 15% of proteins having trans-membrane domain in the case of *A. thaliana* cell walls. Unless looking for specific proteins only extractible in those conditions, this step should be avoided for a large-scale cell wall proteomic study. It can rather be a good method to get rid of contaminants and to have a cleaner preparation for subsequent extraction of strongly bound CWP.

## Conclusion

The new cell wall preparation procedure followed by salt extraction of proteins described in this paper gives the lowest proportion of proteins predicted to be intracellular when compared with other available protocols and allows the identification of proteins fitting in the same functional classes. Addition of a step including detergent treatment revealed the presence of minor amounts of a few additional proteins predicted to be secreted, but of many proteins predicted to be intracellular. Prediction of the sub-cellular localization of proteins by different bioinformatic programs appeared as an essential tool to evaluate cell wall purification procedures. However, it should not be considered satisfactory to determine the sub-cellular localization of any protein identified by a proteomic analysis. Additional experiments performed *in planta*, such as immunolocalization or localization of fluorescent protein fusions, are required to confirm it. The application of the principles of cell wall purification described in this paper should lead to a more realistic view of the cell wall proteome, at least for the weakly bound CWP extractable by salts. In addition, it offers a clean cell wall preparation for subsequent extraction of strongly bound CWP.

## Methods

### *Plant material and isolation of cell walls*

One hundred and fifty mg of *A. thaliana* seeds (ecotype Columbia 0) were sowed per Magenta box containing Murashige and Skoog medium [26] supplemented with 2% w/v sucrose and 1.2% w/v agar. Seedlings were grown at 23°C in the dark for 11 days. For one experiment, hypocotyls from 16 Magenta boxes were collected. One cm long hypocotyls were cut below the cotyledons and above the root, washed with distilled water and transferred into 500 mL of 5 mM acetate buffer, pH 4.6, 0.4 M sucrose and protease inhibitor cocktail (Sigma) 1 mL per 30 g of hypocotyl fresh weight. The mixture was ground in a blender at full speed for 15 min (Figure 4). After adding PVPP (1 g per 10 g fresh weight of hypocotyls), the mixture was incubated in cold room for 30 min while stirring. Cell walls were separated from soluble cytoplasmic fluid by centrifugation of the homogenate for 15 min at 1000 × g and 4°C. The pellet (CW1 in Figure 4) was further purified by two successive centrifugations in 500 mL of 5 mM acetate buffer, pH 4.6, respectively 0.6 M and 1 M sucrose. The residue (CW3) was washed with 3 L of 5 mM acetate buffer, pH 4.6, on a nylon net (25 µm pore size). The resulting cell wall fraction (CW4) was ground in liquid nitrogen in a mortar with a pestle prior to lyophilization. Starting with 16 g fresh weight of hypocotyls, this process resulted in 1.3 g dry powder.

### *Sequential proteins extraction and identification*

Typically, 0.65 g of lyophilized cell walls was used for one experiment. Proteins were extracted by successive salt solutions in this order: two extractions each time with 6 mL $CaCl_2$ solution (5 mM acetate buffer, pH 4.6, 0.2 M $CaCl_2$ and 10 µL protease inhibitor cocktail), followed by two extractions with 6 mL LiCl solution (5 mM acetate buffer, pH 4.6, 2 M LiCl and 10 µL protease inhibitor cocktail). Cell walls were resuspended by vortexing for 5–10 min at room temperature, and then centrifuged for 15 min at 4000 × g and 4°C. Supernatants were desalted using Econo-Pac® 10 DG columns (Bio-Rad) equilibrated with 0.2 formic acid ammonium salt. The extract were lyophilized and resuspended in sample buffer for separation of proteins by 1D-GE, as previously described [12].

The next extraction was carried out by SDS and DTT. The cell wall preparation was treated with 12 mL solution containing 62.5 mM Tris, 4% SDS, 50 mM DTT, pH 6.8 (HCl). The mixture was boiled for 5 min and centrifuged for 15 min at 40000 × g and 4°C. The supernatant was dialyzed against 1 L $H_2O$ in Spectra/Por® membrane 10 kDa MWCO bags (Spectrum Medical Industries) at room temperature, then concentrated by successive centrifugation using the Centriprep® centrifugal filter devices (YM-10 kDa membrane) (Millipore) at 4000 × g followed by speed vacuum centrifugation.

The protein content of each extract was measured using the Bradford method [27] with the Coomassie™ protein assay reagent kit (Pierce) using bovine serum albumin (BSA) as standard.

Gels were stained with Coomassie™ Brilliant Blue-based method [28]. Colored bands were digested with trypsin and MALDI-TOF MS or LC-MS/MS analyses were performed as previously reported [12,13].

The sequences of the identified proteins were subsequently analyzed with several bioinformatic programs to predict their sub-cellular localization [29-31]. In some cases, predictions were not the same with the three programs. Results are then indicated as "not clear". Data are described in Tables 1–5 (additional data).

### *Abbreviations*

1D/2D-GE: one or two dimensional gel electrophoresis

CWP: cell wall proteins

CHAPS: 3- [(3-cholamidopropyl)dimethylammonio]-1-propanesulfonic acid

DTT: dithiothreitol

EDTA: ethylene diamino tetraacetic acid

LC-MS/MS: liquid chromatography-tandem mass spectrometry

MALDI-TOF: matrix-assisted laser desorption ionization-time of flight

MS: mass spectrometry

PMSF: phenylmethylsulfonyl fluoride

PVPP: polyvinyl polypyrrolidone

SDS: sodium dodecyl sulphate

## Competing interests

The author(s) declare that they have no competing interests.

## Authors' contributions

LF and MI carried out cell wall preparations, protein extractions, protein separations and mass spectrometry analyses. HC designed the cell wall preparation procedure and contributed to MALDI-TOF analyses and manuscript writing. EJ conceived data evaluation, performed bioinformatic analyses and was involved in manuscript writing. RPL participated in discussions all along the study, in literature screening, conceived the manuscript, and coordinated its writing. All authors read and approved the final manuscript.

## Additional material

### Additional data file 1

***Table 1** – Bioinformatic analysis of proteins extracted from cell walls of* A. thaliana *cell suspension cultures [8].*
Click here for file
[http://www.biomedcentral.com/content/supplementary/1746-4811-2-10-S1.pdf]

### Additional data file 2

***Table 2** - Bioinformatic analysis of proteins extracted from cell walls of* M. sativa *stems [9]*
Click here for file
[http://www.biomedcentral.com/content/supplementary/1746-4811-2-10-S2.pdf]

### Additional data file 3

***Table 3** – Bioinformatic analysis of proteins extracted from cell walls of* C. albicans *[20].*
Click here for file
[http://www.biomedcentral.com/content/supplementary/1746-4811-2-10-S3.pdf]

### Additional data file 4

***Table 4** - Bioinformatic analysis of proteins extracted from cell walls of* A. thaliana *etiolated hypocotyls with salts.*
Click here for file
[http://www.biomedcentral.com/content/supplementary/1746-4811-2-10-S4.pdf]

### Additional data file 5

***Table 5** - Bioinformatic analysis of proteins extracted from cell walls of* A. thaliana *etiolated hypocotyls with SDS and DTT.*
Click here for file
[http://www.biomedcentral.com/content/supplementary/1746-4811-2-10-S5.pdf]

## Acknowledgements

The authors are grateful to the Université Paul Sabatier (Toulouse, France) and the CNRS (France) for support. M.I. is a fellow from the Higher Education Commission, Islamabad, Pakistan and from the French government on the behalf of SFERE. Mass spectrometry analyses were performed on the *Plate-Forme de Spectrométrie de Masse* in Toulouse, France.